\documentstyle[array,epsfig,preprint,pra,aps]{revtex}

\newcommand{\ep}{\epsilon}
\newcommand{\wt}{\widetilde}
\begin{document}
\draft

  
\title{Quantum Nonlocality for a Mixed Entangled Coherent State} 
  
  \author{D. Wilson,
     H. Jeong
     and M. S. Kim
    }
  
  \address{School of Mathematics and Physics, Queen's University,
    Belfast BT7 1NN, United Kingdom 
    }
  
  \date{\today}
  
  \maketitle

\begin{abstract}
  Quantum nonlocality is tested for an entangled coherent state,
  interacting with a dissipative environment.  A pure entangled
  coherent state violates Bell's inequality regardless of its coherent
  amplitude. The higher the initial nonlocality, the more rapidly
  quantum nonlocality is lost.  The entangled coherent state can also
  be investigated in the framework of $2\times2$ Hilbert space.  The
  quantum nonlocality persists longer in $2\times2$ Hilbert space.  
  When it decoheres it is found that the entangled
  coherent state fails the nonlocality test, which contrasts
  with the fact that the decohered entangled state is always
  entangled. 
\end{abstract}



\newpage
\section{INTRODUCTION}
Einstein, Podolsky, and Rosen proposed a thought experiment to test the 
local realism properties of quantum mechanics \cite{EPR}.  An
inequality imposed by a local hidden variable theory was suggested by
Bell \cite{Bell}, which enables a quantitative test concerning the
controversy of local realism versus quantum mechanics in a real
laboratory.  Various versions of Bell's inequality \cite{CHSH,CH}
followed the original one, and experiments have been performed testing
local realism \cite{exp}.  In spite of successful experimental results,
it has been pointed out that there remain two possible loopholes.
One is called the lightcone loophole that might allow local realistic
interpretation.  Some experiments have been performed with strict
relativistic separation between measurements to close this loophole
\cite{lh1}.  The other is the detection loophole due to detection
inefficiency.  According to this loophole, there is a possibility that the
detected subensemble violates Bell's inequality even though the whole
ensemble satisfies it.  Some authors generalised Bell's inequality to
the case of inefficient detection \cite{CH,GM,Larsson} and other
proposals have been made \cite{lh2} to close the detection loophole.
An experiment on nonlocality has been performed with an efficient detection 
\cite{Rowe}.

\begin{figure}
\centerline{\scalebox{0.5}{\includegraphics{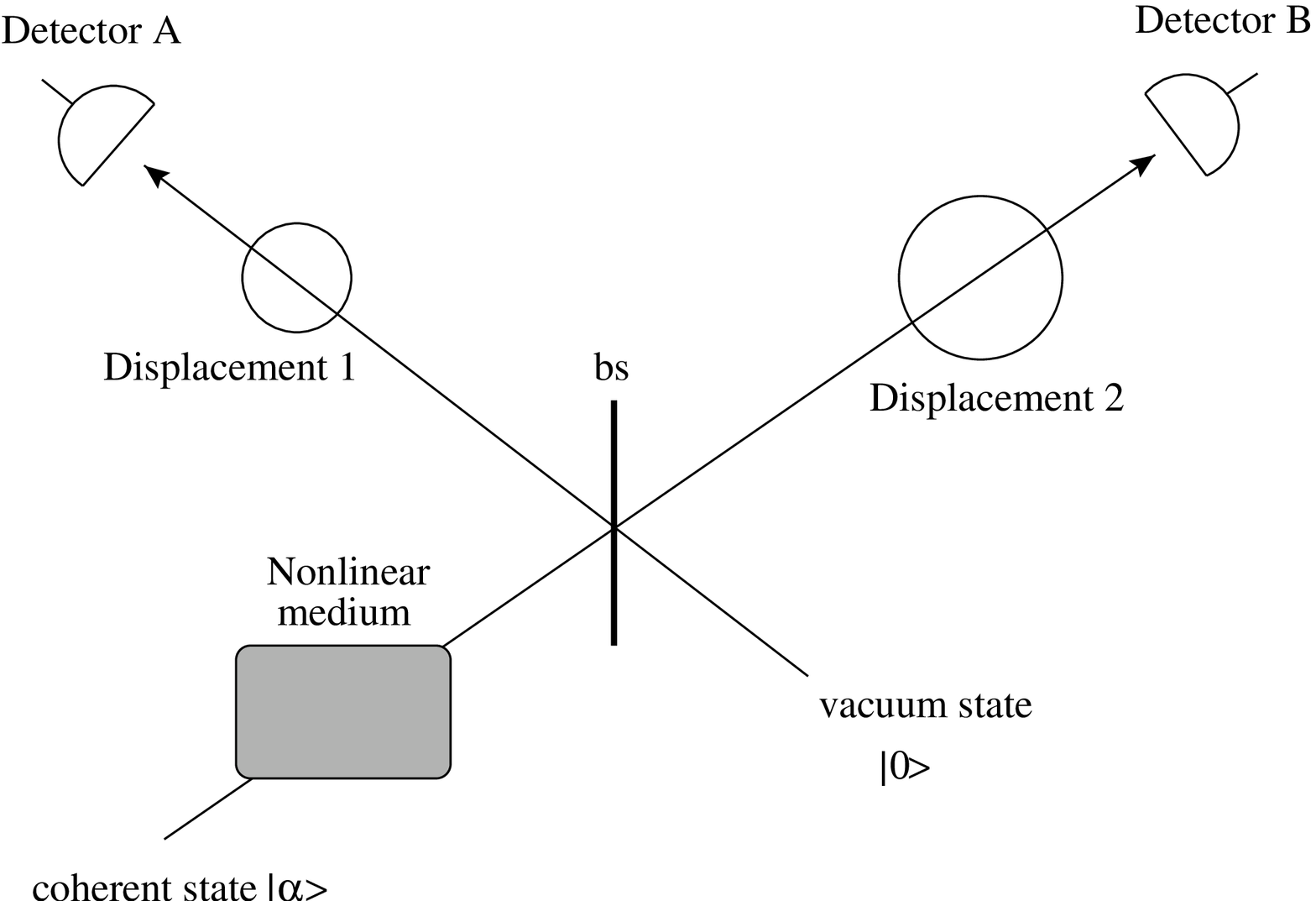}}}
\vspace{0.5cm}
\caption{Nonlocality test for an entangled coherent state.  
  A coherent state, nonlinear
  medium, and 50-50 beam splitter are used to generate an entangled
  coherent state.}
\label{nonlocaltest}
\end{figure}

The nonlocality test can be performed on an entangled system composed of two 
coherent systems \cite{Sanders92}.  This entangled system can be used as a 
quantum entangled 
channel for quantum information transfer. The entangled coherent state can be
generated using a coherent light propagating through a nonlinear
medium \cite{Yurke} and a 50-50 beam splitter as shown in
Fig.~\ref{nonlocaltest}.  There have also been proposals to
entangle fields in two spatially separated cavities
\cite{Davidovich93}.  The entangled coherent state and its usage for
quantum information processing \cite{Enk00,Hirota01,Wang01,JKL01}, 
teleportation \cite{Enk00,Wang01,JKL01} and entanglement concentration 
\cite{JKL01} have all been studied recently.

In this paper, we study nonlocality of an entangled coherent state
using photon parity measurement.  We first investigate the nonlocality of
a pure entangled coherent state, and move on to the dynamic behavior of
nonlocality for a decohered entangled coherent state in a vacuum
environment.  The dynamic behaviour is also
investigated in the framework of $2\times2$ Hilbert space and the
results are compared.  It is found that nonlocality of a decohered  entangled
coherent sate persists longer when it is considered in $2\times2$ space.

\section{Nonlocality for an entangled coherent state}
We are interested in nonlocality of an entangled coherent state
\begin{equation}
\label{eq:ecs1}
|C\rangle_{12}=\frac{1}{\sqrt{N}}(|\alpha\rangle_1|-\alpha\rangle_2+e^{i\varphi} |-\alpha\rangle_1|\alpha\rangle_2),
\end{equation}
where $\alpha=\alpha_r+i\alpha_i$ is a complex amplitude, $\varphi$ is
a real local phase factor, and $N$ is a
normalisation factor.  By applying local unitary transformations any  
state in the form of
\begin{equation}
\label{eq:ecs2}
|\Psi\rangle_{12}=\frac{1}{\sqrt{N}}(|\beta\rangle_1|\gamma\rangle_2+e^{i\varphi} |\gamma\rangle_1|\beta\rangle_2)
\end{equation}
with arbitrary amplitudes $\beta$ and $\gamma$ can be transformed to a
form like that of Eq.~(\ref{eq:ecs1}) up to a global phase factor. 
Applying displacement operators $D_1(x)D_2(x)$, where $x=x_r+i x_i$ is complex,
the entangled coherent state (\ref{eq:ecs2}) becomes
\begin{equation}
D_1(x)D_2(x)|\Psi\rangle_{12}
=\frac{e^{i\phi}}{\sqrt{N}}(|\alpha\rangle_1|-\alpha\rangle_2+e^{i\varphi} |-\alpha\rangle_1|\alpha\rangle_2)
\end{equation}
with 
$x=-\frac{1}{2}(\beta+\gamma)$, $\alpha=\beta+x=-(\gamma+x)$, and $\phi=x_i
\beta_r-x_r \beta_i+x_i\alpha_r-x_r \alpha_i$.  Here the displacement operator is
defined as $D(x)=\exp(xa^\dag+x^* a)$, where $a$ and $a^\dag$ are annihilation and creation operators \cite{Cahill}.   
In this paper, we consider the nonlocality of entangled coherent states
\begin{equation}
\label{c+-}
|C_\pm\rangle={1 \over \sqrt{N_\pm}}(|\alpha\rangle|-\alpha\rangle\pm|-\alpha\rangle|\alpha\rangle)
\end{equation}
where $N_\pm$ are normalisation factors and $\alpha$ is assumed to be real for simplicity. 

The type of quantum observable to be measured is a crucial factor in the
nonlocality test.  Banaszek and W{\' o}dkiewicz developed a Wigner function
representation of the Bell-Clauser, Horne, Shimony and Holt (CHSH) \cite{CHSH} inequality using a two-mode parity
operator
$\Pi(\alpha,\beta)$ as a quantum observable \cite{BW98,BW99}.  
The two-mode parity operator $\Pi(\alpha,\beta)$ is defined as
\begin{equation}
\label{eq:pm1}
\Pi(\alpha,\beta)=D_1(\alpha)D_2(\beta)\Pi D_1^\dagger(\alpha)D_2^\dagger(\beta),
\end{equation}
where 
\begin{equation}
\label{parity-1}
\Pi= \Pi_{e1}\otimes\Pi_{e2}-\Pi_{e1}\otimes\Pi_{o2}-\Pi_{o1}\otimes\Pi_{e2}+\Pi_{o1}\otimes\Pi_{o2},\label{eq:pm2}\\
\end{equation}
with 
\begin{equation}
\label{parity-2}
\Pi_e=\sum_n^\infty|2n\rangle\langle 2n|,~~
\Pi_o=\sum_n^\infty|2n+1\rangle\langle 2n+1|.
\label{eq:pm3}
\end{equation}
The Bell-CHSH inequality is then
\begin{equation}
\label{Bell-CHSH-1}
|{\cal B}|=|\langle\Pi(\alpha,\beta)+\Pi(\alpha,\beta^\prime)+\Pi(\alpha^\prime,\beta)-\Pi(\alpha^\prime,\beta^\prime)\rangle|\leq2,
\end{equation}
where we call $|{\cal B}|$ the Bell measure.
The displacement operation can be effectively performed using a beam
splitter with the transmission coefficient close to one and a strong
coherent state being injected into the other input port \cite{BW99}.
The two-mode Wigner
function at a given phase point described by $\alpha$ and $\beta$ is
\begin{equation}
\label{Wigner-parity}
W(\alpha,\beta)=\frac{4}{\pi^2}{\rm Tr}[\rho\Pi(\alpha,\beta)],
\end{equation}
where $\rho$ is the density operator of the field.
From Eqs.(\ref{Bell-CHSH-1}) and (\ref{Wigner-parity}), we obtain the Wigner representation of Bell's inequality
\begin{equation}
\label{eq:bell}
|{\cal B}|=\frac{\pi^2}{4}|W(\alpha,\beta)+W(\alpha,\beta^\prime)+W(\alpha^\prime,\beta)-W(\alpha^\prime,\beta^\prime)|\leq2.
\end{equation}

\begin{figure}
\centerline{\scalebox{1}{\includegraphics{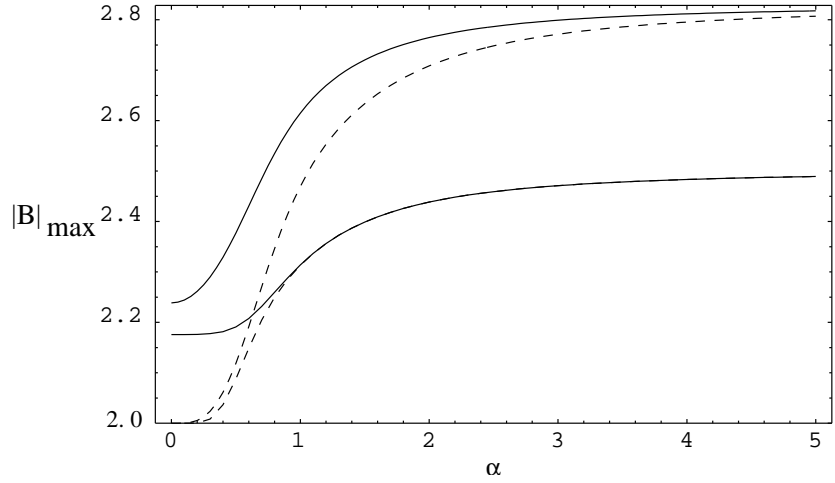}}}
\vspace{0.5cm}
\caption{Bell measure against amplitude $\alpha$ ($>0$), of $|C_-\rangle$ (solid lines)
  and $|C_+\rangle$ (dashed lines) entangled coherent states.  
  The higher valued solid and dashed lines are for
  the generalised Bell measures while the lower valued solid and dashed lines are 
  for the case taking $\alpha=\beta=0$.}
\label{nonlBWcompfull4alpha0to5}
\end{figure}

The Wigner function of an entangled coherent state is obtained from the Fourier
transform of it's characteristic function
\begin{equation}
\label{characteristic-Weyl}
C(\eta,\xi)={\rm Tr}[\rho D_1(\eta)D_2(\xi)].
\end{equation}
Banaszek and W{\'o}dkiewicz used their Wigner function based Bell measure with
$\alpha=\beta=0$.  We consider all four variables $\alpha, \beta, \alpha^\prime$ 
and $\beta^\prime$ in our investigation of the 
Wigner function based Bell measure as shown in Eq.~(\ref{eq:bell}) to test nonlocality more generally.
The maximum Bell-CHSH violation using Banaszek and W{\'o}dkiewicz's version is approximately 2.19 for 
two-mode squeezed states \cite{BW98,JLK00} and 2.5 for entangled coherent states 
as shown in Fig.~\ref{nonlBWcompfull4alpha0to5}.  However, for the  
generalised Bell measure in Eq.~(\ref{eq:bell}), both two-mode squeezed 
states \cite{BSC80} and entangled coherent states have a maximal Bell-CHSH violation of 
$2\sqrt{2}$ (as shown in Fig.~\ref{nonlBWcompfull4alpha0to5}). 

The entangled coherent states $|C_-\rangle$ and $|C_+\rangle$ violate the Bell
inequality regardless of the size of the amplitude, $\alpha>0$. As the amplitude
$\alpha$ increases the Bell measure tends towards a
maximum of $2\sqrt{2}$.

It is interesting to note that the Bell measure for the 
$|C_-\rangle$ state takes a higher value than for the $|C_+\rangle$ state.  If the number state 
representation of the coherent state is \cite{Cahill}
\begin{equation}
|\alpha\rangle = \sum_n \frac{e^{-|\alpha|^2/2}\alpha^n}{\sqrt{n!}}|n\rangle
\end{equation}
which means that, in the limit of small coherent amplitude $\alpha$,
\begin{equation}
|\pm\alpha\rangle \approx e^{-|\alpha|^2/2}[|0\rangle \pm \alpha|1\rangle].
\end{equation}
Substituting this into the entangled coherent states $|C_+\rangle$ and 
$|C_-\rangle$ we find
\begin{eqnarray}
|C_+\rangle &\propto & |0\rangle|0\rangle - \alpha^2|1\rangle|1\rangle
\nonumber \\
|C_-\rangle &\propto& \alpha(|1\rangle|0\rangle - |0\rangle|1\rangle)
\end{eqnarray}
When $\alpha$ is small $|C_+\rangle \rightarrow |0\rangle|0\rangle$.  As the weights of
$|0\rangle|0\rangle$ and $|1\rangle|1\rangle$ are radically different 
$|C_+\rangle$ is only minimally entangled.  However the two component states
$|0\rangle|1\rangle$ and $|1\rangle|0\rangle$ are equally weighted for 
$|C_-\rangle$ which gives optimal entanglement.
A pure entangled state always violates nonlocality \cite{Gisin91}.
Any entangled coherent state in a form as given in 
Eq.~(\ref{c+-}) is found to be nonlocal.  We conjecture that any  
entangled coherent state as given in Eq.~(\ref{eq:ecs2}) also has nonlocality 
except when $\beta=\gamma$, {\em i.e.}, when the state is a product state.

The results shown in Fig.~\ref{nonlBWcompfull4alpha0to5} were obtained from a
numerical consideration of the Bell measure (\ref{eq:bell}).  In analogy with 
the work carried out in \cite{JLK00} we imposed the condition  
$\cal{B}(|\alpha|,|\beta|,|\alpha'|,|\beta'|) = 
\cal{B}(|\beta|,|\alpha|,|\beta'|,|\alpha'|)$. The method of steepest descent \cite{methsteepdescnumrec} 
was used to find the maximum of the Bell measure under our assumptions.

\section{Dynamics of Nonlocality}
A quantum system loses its quantum characteristics if it is open to
the world.  We modelled an entangled coherent state interacting with a
dissipative environment (two independent vacuum reservoirs).  To study the
dynamics of nonlocality of continuous variable entangled coherent states it is
necessary to find an expression of the time-dependent Bell-CHSH measure.  This
in turn means finding an expression for the time-dependent decohered Wigner
function.

The quantum channel decoheres when it interacts with it's environment and 
becomes a mixed state of its density operator $\rho(\tau)$, where $\tau$ is the 
decoherence time. To know the time dependence of $\rho(\tau)$, we have to solve 
the master equation 
\begin{equation}
\label{master-eq}
{\partial \rho \over \partial \tau}=\hat{J}\rho +\hat{L}\rho~;~\hat{J}\rho=\gamma \sum_i a_i\rho a_i^\dag,~~
\hat{L}\rho=-{\gamma \over 2}\sum_i(a_i^\dag a_i\rho +\rho a_i^\dag a_i)
\end{equation}
where $a_i$ and $a_i^\dag$ are the annihilation and creation operators for the 
field mode $i$ and $\gamma$ is the decay constant.  We have assumed that each field mode is coupled to
its environment at the same coupling rate $\gamma$. The formal solution of the master 
equation (\ref{master-eq}) can be written as 
\begin{equation}
\label{formal-sol}
\rho(t)=\exp[(\hat{J}+\hat{L})\tau]\rho(0).
\end{equation}
which leads to the solution for the initial single-mode dyadic $|\alpha\rangle\langle\beta|$
\begin{equation}
\label{solution-master}
\exp[(\hat{J}+\hat{L})\tau]|\alpha\rangle\langle\beta|=\langle\beta|\alpha\rangle^{1-t^2}
|\alpha t\rangle\langle\beta t|
\end{equation}
where $t=e^{-\frac{1}{2}\gamma\tau}$.  In this paper, we introduce a dimensionless normalised 
interaction time $r$ which is related to $t$ by the expression $r=\sqrt{1-t^2}$.
When $\tau=0, t=1$ and $r=0$. As $\tau\rightarrow \infty, t\rightarrow 0$ and 
$r\rightarrow 1$.

\begin{figure}
\centerline{\scalebox{1}{\includegraphics{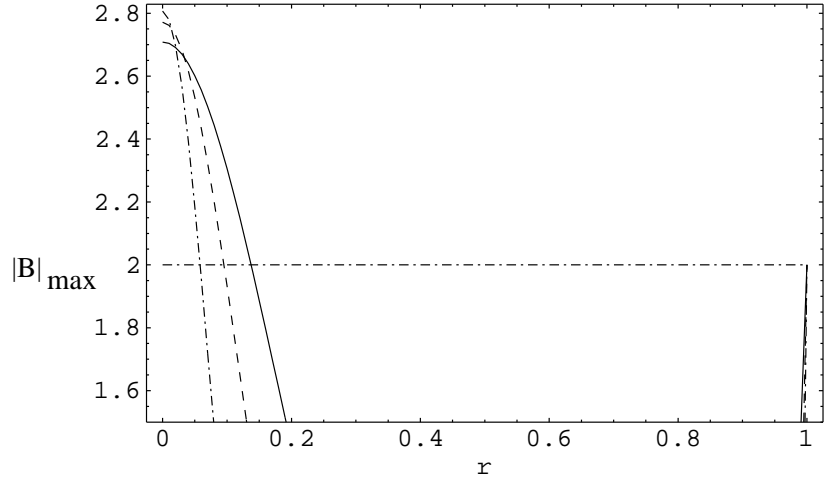}}}
\vspace{0.5cm}
\caption{Nonlocality as a function of the dimensionless normalised time $r$ for
the $|C_-\rangle$ state in the vacuum.  $\alpha=2$ 
(solid line), $\alpha=3$ (dashed line) and $\alpha=5$ (dot-dashed line).}
\label{nonlnegnbar0AR235comp}
\end{figure}

\begin{figure}
\centerline{\scalebox{1}{\includegraphics{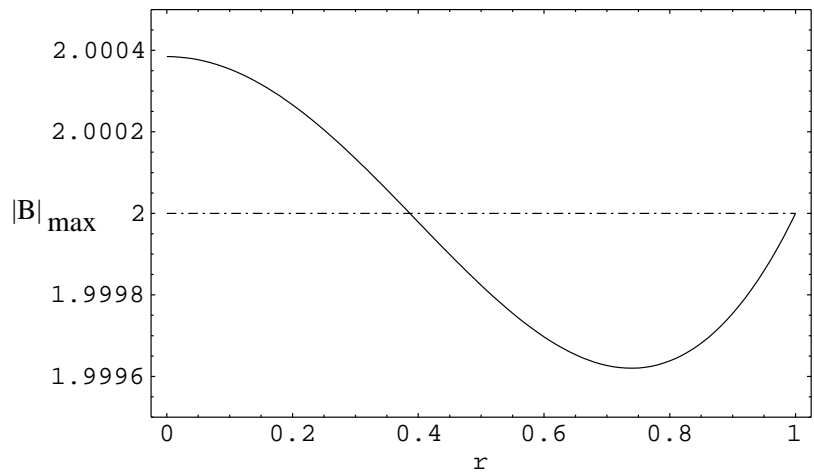}}}
\vspace{0.5cm}
\caption{The $|C_+\rangle$ state for  the coherent amplitude $\alpha=0.1$,
coupled to the vacuum environment, produces a prolonged nonlocal state.}
\label{nonlposnbar0AR01zoom}
\end{figure}

After solving the master equation (\ref{master-eq}) for the initial entangled coherent state, the time-dependent density
operator $\rho(\tau)$ is obtained.  Substituting $\rho(\tau)$ into Eq.(\ref{characteristic-Weyl}),
we calculate the characteristic function and its Fourier transform to obtain the Wigner function for the
decohered entangled coherent state.
Once again the results shown in Figs.~\ref{nonlnegnbar0AR235comp}-\ref{nonlposnbar0AR01zoom}
 were obtained using the method of steepest descent
to find the maximum value of the Bell measure under our assumptions.  The same 
symmetrical consideration as before (namely 
$\cal{B}(|\alpha|,|\beta|,|\alpha'|,|\beta'|)
 = \cal{B}(|\beta|,|\alpha|,|\beta'|,|\alpha'|)$) was imposed.

From Figs.~\ref{nonlnegnbar0AR235comp}-\ref{nonlposnbar0AR01zoom} it is 
obvious that as the entangled coherent state interacts with its environment it 
fails the nonlocality test.  From Fig. \ref{nonlnegnbar0AR235comp} it can be seen 
that as the coherent amplitude $\alpha$ increases the initial nonlocality 
increases and the rate of loss of nonlocality increases.   The 
larger the initial amplitude, {\em i.e.}, the larger the initial nonlocality, the 
more rapid the loss of nonlocality occurs, {\em i.e.}, the shorter the duration of 
the nonlocality.  As $r\rightarrow 1$, in Fig.~\ref{nonlnegnbar0AR235comp}, 
$\rho$ becomes a product of two vacuum states and the Bell measure approaches 
the value 2.  

We can see in Fig.~\ref{nonlposnbar0AR01zoom} that the 
$|C_+\rangle$ state of the coherent amplitude $\alpha=0.1$ has a long duration of nonlocality
($r\approx0.375$). The duration 
of the nonlocality can be increased by decreasing the coherent amplitude.

\section{Nonlocality test in 2$\times$2 dimensional Hilbert space}
Entangled coherent states can be considered in $2\times2$ dimensional
Hilbert space \cite{JKL01}, where $|C_-\rangle$ shows maximal
entanglement regardless of the value of $\alpha$ \cite{Hirota01}.  In this 
section, we will investigate nonlocality and the dynamics of the entangled
coherent state $|C_-\rangle$ in a vacuum environment within the
framework of $2\times2$ Hilbert space (see also the discussions in ref. \cite{Munro-Milburn}).

We consider two orthogonal states
\begin{eqnarray}
\label{eq:basis1}
|e\rangle&=&\frac{1}{\sqrt{{\cal N}_+}}\Big(|\alpha\rangle+|-\alpha\rangle\Big), \\
\label{eq:basis2}
|d\rangle&=&\frac{1}{\sqrt{{\cal N}_-}}\Big(|\alpha\rangle-|-\alpha\rangle\Big)
\end{eqnarray}
where ${\cal N}_+=2+2e^{-2\alpha^2}$ and ${\cal N}_-=2-2e^{-2\alpha^2}$ are
normalisation factors. A two-dimensional Hilbert space can be spanned
using these states as orthonormal bases. The entangled coherent state $|C_-\rangle$ can 
be represented in 2$\times$2 dimensional Hilbert space as
\begin{equation}
|C_-\rangle_{12}=\frac {1} {\sqrt{2}} \Big( |e\rangle_1|d\rangle_2
-|d\rangle_1|e\rangle_2 \Big),
\end{equation}
where we recognise that $|C_-\rangle$ is maximally entangled.

The Bell-CHSH inequality for a bipartite spin-$\frac{1}{2}$ state $|\psi\rangle$
is $|{\cal B}|\leq2$,  where
\begin{equation}
\label{eq:s1}
{\cal B}=\langle\psi|\vec a\cdot \vec\sigma_1\otimes \vec b\cdot \vec\sigma_2  + 
\vec a\cdot \vec\sigma_1\otimes \vec b^\prime\cdot\vec \sigma_2 +  \vec a^\prime\cdot
\vec\sigma_1\otimes \vec b\cdot \vec\sigma_2-   \vec a^\prime\cdot \vec\sigma_1 \otimes\vec
b^\prime\cdot\vec \sigma_2|\psi\rangle\label{eq:s2}
\end{equation}
and $\vec a$, $\vec a^\prime$, $\vec b$ and $\vec b^\prime$ are
three-dimensional unit vectors and $\sigma$'s are Pauli matrices \cite{Br92}.  
The unit vectors determine the directions of $\sigma$-operators which are
measurement observables.  They are usually realised by rotating the measurement 
apparatuses at both sides.
The effect of these unit vectors can also be realised by local unitary
operations on both particles of the pair independently, fixing the
direction of the measurement apparatuses so that the measurement operator
becomes $\sigma_{z1}\otimes\sigma_{z2}$.

We first consider ideal conditions for the nonlocality test. Assume 
$|e\rangle$ and $|d\rangle$ can be perfectly discriminated with
eigenvalues 1 and -1 by an ideal measurement operator
$O_s=|e\rangle\langle e|-|d\rangle\langle d|$, where the operator $O_{s}$ is 
an analogy to $\sigma_z$ in a 
spin-$\frac{1}{2}$ system. 
If an ideal rotation such as $R_x(\theta)$ around an axis,
\begin{eqnarray}
  R_x(\theta)|e\rangle&=&\cos\theta|e\rangle+i\sin\theta|d\rangle,
  \nonumber \\ 
\label{eq:uo}
R_x(\theta)|d\rangle&=&i\sin\theta|e\rangle+\cos\theta|d\rangle,
\end{eqnarray}
can be performed on the particles of both sides by two local measurements
$O_{s1}$ and $O_{s2}$.    Under these conditions, it can be proved that the 
entangled coherent state $|C_-\rangle$ maximally violates the Bell-CHSH 
inequality regardless of the value of $\alpha$, {\em i.e.}, 
$|{\cal B}|_{max}=2\sqrt{2}$.

The dynamic change of nonlocality for the entangled coherent state can be 
obtained from its time-dependent density matrix.  Assuming vacuum environment, 
it is possible to restrict our discussion in $2\times 2$ dimensional Hilbert 
space even for the mixed case. The basis vectors in Eqs.~(\ref{eq:basis1}) and
(\ref{eq:basis2}) now should be
\begin{eqnarray}
\label{eq:decaybasis1}
|e(\tau)\rangle&=&\frac{1}{\sqrt{ {\cal N}_+(\tau)}}( |t\alpha\rangle+|-t\alpha\rangle),~~  \\
\label{eq:decaybasis2}
|d(\tau)\rangle&=&\frac{1}{\sqrt{{\cal N}_-(\tau)}}(|t\alpha\rangle-|-t\alpha\rangle),
\end{eqnarray}
where ${\cal N}_+(\tau)=2+2e^{-2 t^2\alpha^2}$ and ${\cal N}_-(\tau)=2-2e^{-2 t^2\alpha^2}$.
Although $|e(\tau)\rangle$ and $|d(\tau)\rangle$ are time-dependent, they
always remain orthogonal until, $\tau\rightarrow\infty$.

With use of the master equation (\ref{master-eq})
we find the mixed density matrix $\rho_-(\tau)$ as follows
\begin{equation}
\rho_-(\tau)=\frac{1}{4 {\cal N}_+ {\cal N}_-}\left(
\begin{array}{cccc}
A & 0 & 0 & D \\
0 & C & -C & 0 \\
0 & -C & C & 0  \\
D & 0 & 0 & E 
\end{array}\right) \label{eq:Tmatrix},
\end{equation}
where $A$, $C$, $D$ and $E$ are defined as
\begin{eqnarray}
A&=&(1-\Gamma){\cal N}_+^2(\tau),\\
C&=&(1+\Gamma){\cal N}_+(\tau){\cal N}_-(\tau),\\
D&=&-(1-\Gamma){\cal N}_+(\tau){\cal N}_-(\tau),\\
E&=&(1-\Gamma){\cal N}_-^2(\tau),\\
\Gamma&=& \exp\{-4(1-t^2)\alpha^2\}.
\end{eqnarray}

The maximal Bell-CHSH violation for a $2\times2$ dimensional state $\rho$ is 
given in \cite{Horodecki95}
\begin{equation}
|{\cal B}|_{max}=2\sqrt{M(\rho)},
\end{equation}
where $M(\rho)$ is the sum of the two larger eigenvalues of $T
T^\dagger$ and $T$ is a $3\times3$ matrix whose elements are defined
as $t_{nm}={\mathrm Tr}(\rho\sigma_m\otimes\sigma_n)$ with Pauli
matrices represented by $\sigma's$.  Ideal measurement and rotation ability 
should be assumed again here to use this formula.  The three eigenvalues of $T
T^\dagger$ for the mixed entangled coherent state $\rho_-$ are
\begin{equation}
\frac{(C+D)^2}{4 {\cal N}_+^2 {\cal N}_-^2},~ \frac{(C-D)^2}{4 {\cal
 N}_+^2 {\cal N}_-^2},~ \frac{(A-2C+E)^2}{16 {\cal N}_+^2 {\cal N}_-^2},
\end{equation}
from which $M(\rho_-)$ is obtained by calculating the sum of the two larger
eigenvalues.

\begin{figure}
\centerline{\scalebox{1}{\includegraphics{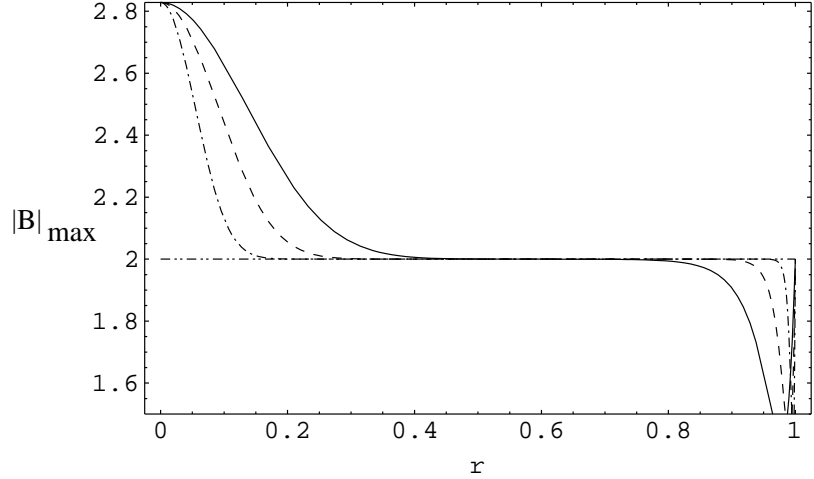}}}
\vspace{0.5cm}
\caption{Bell measure for an entangled coherent state against
  normalised time $r$ in $2\times2$ Hilbert space under perfect rotations.
  Nonlocality persists longer in $2\times2$ space than in continuous
  Hilbert space. $\alpha=2$ (solid line), $\alpha=3$ (dashed line) and
  $\alpha=5$ (dot-dashed line).}
\label{fig:b}
\end{figure}

Fig.~\ref{fig:b} shows $|{\cal B}|_{max}$ versus the dimensionless
time $r(\tau)$. Initially, $\rho_-(\tau=0)$ is maximally entangled
regardless of $\alpha$ and $|{\cal B}|_{max}$ has the maximal value
$2\sqrt{2}$.  As interaction time increases, nonlocality decreases.  
For $\tau\rightarrow\infty$, $\rho_-$ becomes a direct product of two 
coherent states, which is a pure state, and $|{\cal B}|_{max}$ becomes 2.  
It is clear from Fig.~\ref{fig:b} that the nonlocality persists longer in 
$2\times2$  Hilbert space than in continuous-variable space.  We can see that 
the nonlocality of a given state varies according to the Hilbert space in 
which the state is considered, as entanglement also does \cite{JKL01}.

It has already been found that the decohered state $\rho_-(\tau)$ always
remains entangled in $2\times2$ Hilbert space \cite{JKL01}.  This
indicates that the mixed state $\rho_-(\tau)$ retains some amount of
entanglement even after it loses its nonlocality.  For pure states, it
is true that any entangled state violates Bell's inequality
\cite{Gisin91}.  On the other hand, it was shown that there are mixed
states which are entangled but do not violate Bell's inequality
\cite{PH}.  Our model in $2\times2$ space is one example of
that case.

Because the state $|e\rangle$ contains only even numbers of photons and
$|d\rangle$ contains only odd numbers of photons, these two states are
eigenstates of the operator $O_r=\Pi_e-\Pi_o$ which is known as the
pseudo-spin operator \cite{Chen01}, {\em i.e.},
\begin{eqnarray}
&&O_r|x_n\rangle=\lambda_n|x_n\rangle;~~~n=1,2 \\
&&\lambda_{1,2}=\pm1;~~~|x_{1,2}\rangle=|e\rangle,|d\rangle,
\end{eqnarray}
by which $|e\rangle$ and $|d\rangle$ can be perfectly discriminated.
The parameters $\lambda_{1,2}$ are eigenvalues of the pseudo-spin operator $O_r$ and
$|x_{1,2}\rangle$ are eigenvectors of the operator.  The measurement
for the nonlocality test is now $\Pi=O_{r1}\otimes O_{r2}$, which is in
fact the same as the $\Pi$ defined in Eqs.~(\ref{eq:pm2}) and
(\ref{eq:pm3}).  Therefore, the nonlocality test in $2\times2$ space can
be performed by the same parity measurement as in Eqs.~(\ref{parity-1}) and (\ref{parity-2}).  Note that there is no
way to distinguish between $O_r$ and $O_s$ in our restricted Hilbert space.

If an ideal rotation $R_x$ is possible for $|e\rangle$ and
$|d\rangle$, the same structure as $\vec a\cdot\vec\sigma_1\otimes
\vec b\cdot \vec\sigma_2$ can be perfectly made by $\Pi$. 
Cochrane {\em et al.} \cite{C&M} showed that rotation $R_x(\theta)$ can be approximately realised for $\alpha\gg1$
by a displacement operator 
which can change the parity of the even state $|e\rangle$ and the odd
state $|d\rangle$ \cite{C&M}.  When a displacement operator $D(i\ep)$,
where $\ep$ is real, is applied to a given parity eigenstate it shows
oscillations between $|e\rangle$ and $|d\rangle$.

To obtain the Bell function, we can calculate
\begin{eqnarray}
\label{eq:Pe}
P_e(\ep)&=&\langle e^\prime|\Pi_e|e^\prime\rangle=\frac{ e^{\alpha^2-\ep^2}\Big\{e^{2
    i\alpha\ep}\cosh[(\alpha+i\ep)^2]+ e^{-2
    i\alpha\ep}\cosh[(\alpha-i\ep)^2]+ 2 \cosh[\alpha^2+\ep^2]
\Big\}} {2(1+ e^{2\alpha^2})}, \\
\widetilde{P}_e(\ep)&=&\langle d^\prime|\Pi_e|d^\prime\rangle=\frac{ e^{\alpha^2-\ep^2}\Big\{e^{2
    i\alpha\ep}\cosh[(\alpha+i\ep)^2]+ e^{-2
    i\alpha\ep}\cosh[(\alpha-i\ep)^2]- 2 \cosh[\alpha^2+\ep^2]
\Big\}} {2(1- e^{2\alpha^2})}, \\
I_e(\ep)&=&\langle e^\prime|\Pi_e|d^\prime\rangle=\frac{ e^{\alpha^2-\ep^2}\Big\{e^{2
    i\alpha\ep}\cosh[(\alpha+i\ep)^2]- e^{-2
    i\alpha\ep}\cosh[(\alpha-i\ep)^2]
\Big\}} {2\sqrt{1- e^{-4\alpha^2}}},\\
P_o(\ep)&=&\langle e^\prime|\Pi_o|e^\prime\rangle=1-P_e(\ep),\\
\widetilde{P}_o(\ep)&=&\langle d^\prime|\Pi_o|d^\prime\rangle=1-\widetilde{P}_e(\ep),\\
\label{eq:Io}
I_o(\ep)&=&\langle d^\prime|\Pi_o|e^\prime\rangle=-I_o(\ep),
\end{eqnarray}
where $|e^\prime\rangle=D(i\ep)|e\rangle$ and
$|d^\prime\rangle=D(i\ep)|d\rangle$.  

\begin{figure}
\centerline{\scalebox{1}{\includegraphics{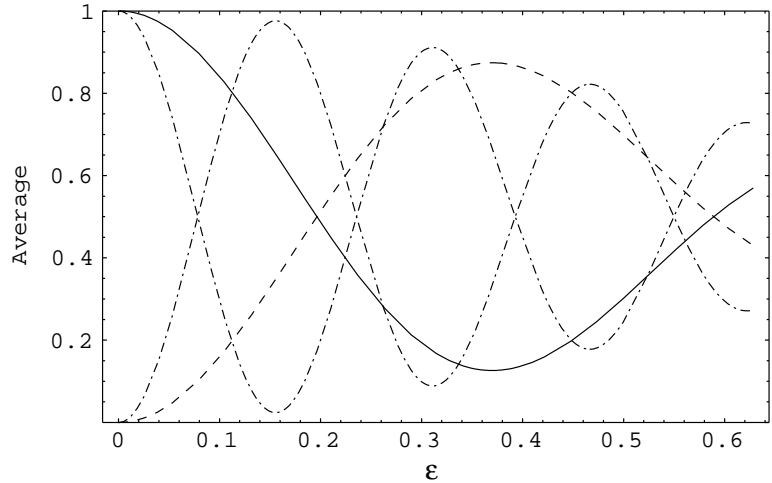}}}
\vspace{0.5cm}
\caption{Oscillations in even and odd states by the displacement
  operator $D(i\ep)$.  For $\alpha\gg1$, the displacement operator acts as a
  sinusoidal rotation. For $\alpha=2$, $\langle e^\prime|\Pi_e|e^\prime\rangle$ (solid line)
and $\langle d^\prime|\Pi_e|d^\prime\rangle$ (dashed line).  
For $\alpha=5$, $\langle e^\prime|\Pi_e|e^\prime\rangle$ (dot-dashed line)
and $\langle d^\prime|\Pi_e|d^\prime\rangle$ (dotted line). }
\label{fig:c}
\end{figure}

From the average values $P_e(\ep)=\langle e^\prime|\Pi_e|e^\prime\rangle$ and
$\widetilde{P}_e(\ep)=\langle d^\prime|\Pi_e|d^\prime\rangle$ shown in
Fig.~\ref{fig:c}, which represent the probabilities for the measured
state to have even parity, we can see oscillations due to 
$D(i\ep)$ in the even and odd states.  

The Bell-CHSH inequality is then obtained using Eqs.~(\ref{eq:Pe}) to 
(\ref{eq:Io}),
\begin{eqnarray}
{\cal B}
&&=\langle C_-(\ep_1,\ep_2)|\Pi| C_-(\ep_1,\ep_2)\rangle+
\langle C_-(\ep_1,\ep_2^\prime)|\Pi| C_-(\ep_1,\ep_2^\prime)\rangle\nonumber\\
&&~~~~~~ +\langle C_-(\ep_1^\prime,\ep_2)|\Pi| C_-(\ep_1^\prime,\ep_2)\rangle-
\langle C_-(\ep_1^\prime,\ep_2^\prime)|\Pi| C_-(\ep_1^\prime,\ep_2^\prime)\rangle\\
\nonumber
&&=\Big(2P_e(\ep_1)-1\Big)\Big(\wt{P}_e(\ep_2)+\wt{P}_e(\ep_2^\prime)-1\Big)+\Big(2\wt{P}_e(\ep_1)-1\Big)\Big(P_e(\ep_2)
+P_e(\ep_2^\prime)-1\Big)\nonumber\\
&&~~~~~+\Big(2P_e(\ep_1^\prime)-1\Big)\Big(\wt{P}_e(\ep_2)-\wt{P}_e(\ep_2^\prime)\Big)+\Big(2\wt{P}_e(\ep_1^\prime)-1\Big)\Big(P_e(\ep_2)-P_e(\ep_2^\prime)\Big)\nonumber\\
&&~~~~~+4I_e(\ep_1)\Big(I_e(\ep_2)+I_e(\ep_2^\prime)\Big)+4I_e(\ep_1^\prime)\Big(I_e(\ep_2)-I_e(\ep_2^\prime)\Big)
,
\end{eqnarray}
where $|C_-(\ep_1,\ep_2)\rangle=D_1(i\ep_1)\otimes D_2(i\ep_2)|C_-\rangle_{12}$.
The nonlocality of this pure entangled coherent state is the same as the
nonlocality of the four variable consideration shown in 
Fig.~\ref{nonlBWcompfull4alpha0to5} (higher valued solid line).

For a mixed state, $\rho_-(\tau)$ is used to obtain the Bell function,
\begin{eqnarray}
&&{\cal B}=
{\rm Tr}\{\rho_-(\tau;\ep_1,\ep_2) \Pi\}
+{\rm Tr}\{\rho_-(\tau;\ep_1,\ep_2^\prime) \Pi\}
+{\rm Tr}\{\rho_-(\tau;\ep_1^\prime,\ep_2) \Pi\}
-{\rm Tr}\{\rho_-(\tau;\ep_1^\prime,\ep_2^\prime) \Pi\},\\
&&\rho_-(\tau;\ep_1,\ep_2)=D_1(i\ep_1)\otimes D_2(i\ep_2)\rho_-(\tau)D_1^\dagger(i\ep_1)\otimes
D_2^\dagger(i\ep_2).
\end{eqnarray}
To calculate ${\rm Tr}\{\rho_-(\tau;\ep_1,\ep_2) \Pi\}$, we need to use
the redefined identity in the restricted Hilbert space, 
\begin{eqnarray}
\openone_r&=&
|e(\tau)\rangle_1|e(\tau)\rangle_{22}\langle e(\tau)|_1\langle e(\tau)|+
|e(\tau)\rangle_1|d(\tau)\rangle_{22}\langle d(\tau)|_1\langle e(\tau)|
\nonumber \\
&&~~~+|d(\tau)\rangle_1|e(\tau)\rangle_{22}\langle e(\tau)|_1\langle d(\tau)|
+|d(\tau)\rangle_1|d(\tau)\rangle_{22}\langle d(\tau)|_1\langle d(\tau)|\nonumber\\
&\equiv&\sum_{n=1}^4|X_n\rangle\langle X_n|,
\end{eqnarray}
which is not equal to the identity $\openone=\frac{1}{\pi}\int
d^2\alpha
d^2\beta|\alpha\rangle|\beta\rangle\langle\beta|\langle\alpha|$ in the
continuous-variable basis.
Using
\begin{equation}
{\rm Tr}\{\rho_-(\tau;\ep_1,\ep_2) \Pi\}=\sum_{n,m}\langle
X_n|\rho_-(\tau)|X_m\rangle\langle X_m| D_1^\dagger(i\ep_1)\otimes
D_2^\dagger(i\ep_2) \Pi D_1(i\ep_1)\otimes D_2(i\ep_2)|X_n\rangle,
\end{equation}
$\cal B$ is obtained by a straightforward calculation as
\begin{eqnarray}
{\cal B}&=&Ave(\ep_1,\ep_2)+Ave(\ep_1,\ep_2^\prime)+Ave(\ep_1^\prime,\ep_2)-Ave(\ep_1^\prime,\ep_2^\prime),\\
Ave(\ep_1,\ep_2)&=&g(\ep_1,\ep_2)A+l(\ep_1,\ep_2)E+2h(\ep_1,\ep_2)(C-D)+\Big(j(\ep_1,\ep_2)+k(\ep_1,\ep_2)\Big)C,\\
g(\ep_1,\ep_2)&=&(2{\bf P}_e(\ep_1)-1)(2\widetilde{\bf P}_e(\ep_2)-1),\\
h(\ep_1,\ep_2)&=& 8{\bf I}_e(\ep_1){\bf I}_e(\ep_2),\\
j(\ep_1,\ep_2)&=&(2{\bf P}_e(\ep_1)-1)(2\widetilde{\bf P}_e(\ep_2)-1),\\
k(\ep_1,\ep_2)&=&(2{\bf P}_e(\ep_1)-1)(2\widetilde{\bf P}_e(\ep_2)-1),\\
l(\ep_1,\ep_2)&=&(2{\bf P}_e(\ep_1)-1)(2\widetilde{\bf P}_e(\ep_2)-1),
\end{eqnarray}
where 
${\bf P}_e(\ep)=\langle e^\prime(\tau)|\Pi_e|e^\prime(\tau)\rangle$, 
$\widetilde{\bf P}_e(\ep)=\langle d^\prime(\tau)|\Pi_e|d^\prime(\tau)\rangle$,
${\bf I}_e(\ep)=\langle e^\prime(\tau)|\Pi_e|d^\prime(\tau)\rangle$,
${\bf P}_o(\ep)=\langle e^\prime(\tau)|\Pi_o|e^\prime(\tau)\rangle=1-{\bf P}_e(\ep)$,
$\widetilde{\bf P}_o(\ep)=\langle d^\prime(\tau)|\Pi_o|d^\prime(\tau)\rangle=1-\widetilde{\bf P}_e(\ep)$,
${\bf I}_o(\ep)=\langle d^\prime(\tau)|\Pi_o|e^\prime(\tau)\rangle=-{\bf I}_o(\ep)$.
These are modified versions of Eqs.~(\ref{eq:Pe}) to (\ref{eq:Io}) with
$|e^\prime(\tau)\rangle=D(i\ep)|e(\tau)\rangle$ and $|d^\prime(\tau)\rangle=D(i\ep)|d(\tau)\rangle$. 
As $\alpha$ increases, it is expected that the result in
$2\times2$ space under the $D(i\ep)$ operation approaches to the ideal
case shown in Fig.~\ref{fig:b}.

\begin{figure}
\centerline{\scalebox{1}{\includegraphics{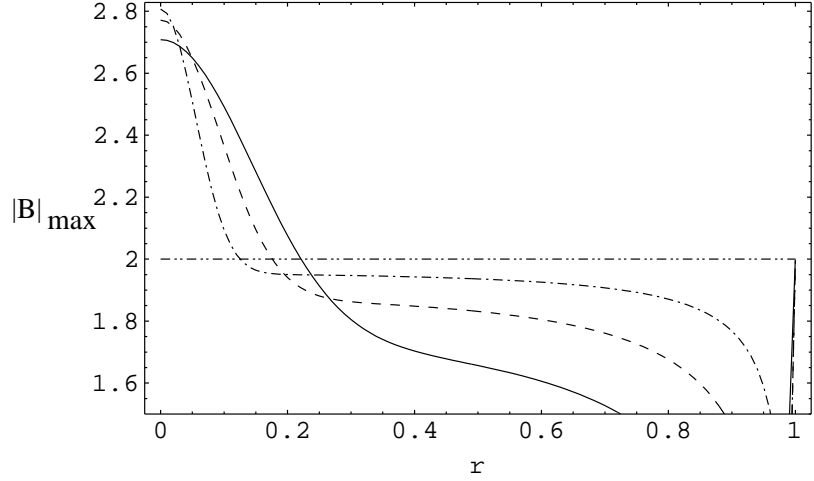}}}
\vspace{0.5cm}
\caption{Bell measure against normalised time for a mixed
  entangled coherent state in $2\times2$ Hilbert space. For
  $\alpha\gg1$, rotation needed for the nonlocality test in the 2-qubit state
  is ideally realised as shown in Fig.~\ref{fig:c}, and the 
  Bell function approaches the ideal case shown in Fig.~\ref{fig:b}.
  $\alpha=2$ (solid line), $\alpha=3$ (dashed line) and $\alpha=5$
  (dot-dashed line). }
\label{fig:e}
\end{figure}

Fig.~\ref{fig:e} shows the largest Bell violation $|{\cal B}|_{max}$ for
$\rho_-(\tau)$ against the normalised time. It is different from the former case of
continuous variable entangled coherent state for $\tau\neq0$, because the 
concerned identities are different from each other.  The nonlocality of a given 
state can differ according to the Hilbert space concerned even though the same 
kind of measurement observable is used.  For $\alpha\gg1$, the rotation
needed for the nonlocality test in the 2-qubit state is ideally realised, and
the time variance of the nonlocality approaches to the ideal case in
Fig.~\ref{fig:b} as was expected.  For $\alpha\ll1$, required
rotation deviates from the ideal case.

\section{Remarks}
We have studied the dynamic behavior of nonlocality for an entangled
coherent state in a dissipative environment.  The nonlocality test for an
entangled coherent state can be realised with photon number
measurement and displacement operations.  Any entangled
coherent state in the form of Eq.~(\ref{eq:ecs2}) can be transformed to a form (\ref{eq:ecs1})
by local unitary
transformations.  The entangled coherent state is
found to be nonlocal regardless of its amplitude.  The
higher the amplitude, the larger the nonlocality is.  When the state interacts
with its environment, the nonlocality is lost.  The rapidity
of the loss of nonlocality depends on the initial amplitude of the
state.  The larger the initial amplitude, {\em i.e.}, the larger the initial
nonlocality, the more rapid the loss of nonlocality occurs.  
An entangled coherent state can be studied in $2\times2$ Hilbert space assuming
a vacuum environment, where the nonlocality of the same state persists
for longer in the dissipative environment.

\acknowledgements

This work has been supported by the UK Engineering and Physical Sciences 
Research Council (GR/R33304). D.W. is grateful for financial support from the
Department of Higher and Further Education Training and Employment (DHFETE) and
the David Bates conference fund. H.J. acknowledges the Overseas Research Student award.

\end{document}